\documentstyle[titlepage]{article}
\begin{document}

\setlength{\baselineskip}{24pt}

\begin{titlepage}

\begin{center}
{\Large INFLUENCE OF THE MEASURE ON SIMPLICIAL QUANTUM GRAVITY
IN FOUR DIMENSIONS} \\[1.5cm]
Wolfgang Beirl, Erwin Gerstenmayer, Harald Markum \\
Institut f\"ur Kernphysik, Technische Universit\"at Wien, A-1040
Vienna, Austria \\[3cm]
\end{center}

\begin{quote}
{\small We investigate the influence of the measure in the path integral
for Euclidean quantum gravity in four dimensions within the Regge calculus.
The action is bounded without additional terms by fixing the average
lattice spacing. We set the length scale by a parameter $\beta$ and
consider a scale invariant and a uniform measure. In the low $\beta$
region we observe a phase with negative curvature and a homogeneous
distribution of the link lengths independent of the measure.  The large
$\beta$ region is characterized by inhomogeneous link lengths distributions
with spikes and positive curvature depending on the measure. \\[1cm]
PACS number(s): 12.25.+e, 04.60.+n}
\end{quote}

\end{titlepage}

General relativity is the profound classical theory of gravitation.
However, a fundamental description of gravity needs a synthesis
with quantum theory. A direct route from classical to quantum
gravity is the sum-over-histories formulation
that considers the Euclidean path integral
\begin{equation}
Z = \int D{\bf g}e^{-I_E({\bf g})}
\end{equation}
as the starting point for investigations of non-perturbative
quantum gravity \cite{hawk,hawking,hartle}. The functional integration extents
over a class of four-geometries ${\bf g}$ with gravitational action
\begin{equation}
-I_E({\bf g})=L_P^{-2} \int d^4x g^{\frac{1}{2}}R,
\end{equation}
where $L_P$ is the Planck length, $R$ the curvature scalar and $g$ the
determinant of the metric ${\bf g}$.
In the following
we consider only geometries with the topology of a four-torus and
therefore the Einstein-Hilbert action is used without surface terms.
The main difficulties of integral (1) are well known.
(i) A unique definition of the measure $D{\bf g}$ does not exist
\cite{menotti}. \linebreak[4]
(ii) The action $I_E$ is not bounded due to fluctuations of the conformal
factor \cite{giddings}.

The aim of this work is to investigate the influence of the measure using
the Regge calculus to approximate the path integral in a systematic
way \cite{hartle,regge}.
For this purpose we perform computer simulations on a simplicial
lattice, $i.e.$ four-simplices which are glued together to form a
piecewise flat four-geometry \cite{berg,hamber,beirl}.

We use a construction derived from a triangulation of the four-torus
which has the advantage that the coordination
numbers of the lattice are close to those of a random lattice \cite{berg}.
Given a Euclidean configuration $\{q_l\}$ where $q_l$ is the squared
length of the link $l$ we can calculate the area $A_t$ of every triangle $t$
and
the deficit angle defined by
\begin{equation}
\delta_t=2\pi - \sum_{s\supset t}\Theta_{s,t}.
\end{equation}
The sum runs over all four-simplices $s$ sharing the same triangle $t$
and $\Theta_{s,t}$ denotes the dihedral angle between the two tetrahedras
in simplex $s$ which have triangle $t$ in common \cite{hartle,regge}.

The Einstein-Hilbert action can be replaced then by a sum over all
triangles
\begin{eqnarray}
-I_E \longrightarrow
 L^{-2}_P \sum_t 2A_t \delta_t =: +L^{-2}_P S_R,
\end{eqnarray}
where the Regge action $S_R$ is defined with a minus sign.
We can now use the path integral of simplicial quantum gravity
to calculate expectation values  within the Regge calculus
\begin{equation}
\langle O\rangle = \frac{ \int D \mu O e^{L^{-2}_P S_R}}
{ \int D \mu e^{L^{-2}_P S_R}},
\end{equation}
where $D\mu$ denotes an integration over different simplicial lattices
\cite{hartle}.
Following Berg \cite{berg} and Hamber \cite{hamber} we hold
the incidence matrices of the lattice fixed and vary the $q_l$,
reducing the integration (5) to a summation over different
configurations $\{q_l\}$.
The difficulties (i) and (ii) appear now within the Regge calculus.

(i) The measure $D\mu$  has to be specified but as in the continuum
case a unique definition does not exist.
In this study we compare the scale invariant measure
\begin{equation}
M1:\quad\quad  D \mu = ( \prod_l \frac{dq_l}{q_l} ) {\cal F} (q_1,...,q_{N_1})
\end{equation}
with the uniform measure
\begin{equation}
M2:\quad\quad  D \mu = ( \prod_l dq_l) {\cal F} (q_1,...,q_{N_1}),
\end{equation}
where $\cal F$ is one for configurations fulfilling the Euclidean triangle
inequalities in four dimensions and zero otherwise.

(ii) The sum $S_R$ in Eq. (4) is not bounded and we distinguish two
types of divergences. The first type is due to a rescaling
$q_l \rightarrow \lambda q_l$ leading to $S_R \rightarrow \lambda S_R$.
The second occurs when some of the four-simplices collapse leading to
triangles $t$ with $\delta_t \rightarrow 2\pi$ and $A_t \rightarrow \infty
$.
The introduction of a cosmological constant term removes only the first
type of divergences \cite{hartle,hamber}. However, for a lattice with a
finite number of
links $N_1$ the action is bounded if we require the average lattice
spacing to stay finite. The prescription
\begin{equation}
\bar{a}^2 = \beta L^{2}_P = N_1^{-1} \sum_l q_l
\end{equation}
defines the average lattice spacing $\bar{a}$ in units of the Planck length
and introduces the coupling parameter $\beta$.

In our study of the scale invariant measure M1 we use the constraint
$\bar{a} = const$
to limit the action.
The incorporation of the measure and the necessary rescaling have been
described by Berg \cite{berg}.

The above constraint (8) implies a cutoff since $q_l < N_1 \beta L_P^2 $
for Euclidean configurations. This suggests for computations with the
uniform measure M2 to impose a cutoff $q_l < const$
for every link.

For system M1 the parameter $\beta$ is the expectation value
$\langle q_l \rangle$ in units of $L_P$. To allow a comparison
of the two measures we computed various expectation values as
a function of $\langle q_l\rangle$ on lattices with $4^4$ vertices.

In Fig. 1 the action density $\langle S_R\rangle/\langle V\rangle$ is
depicted versus $\langle q_l\rangle^{\frac{1}{2}}$ where $V$ is the
total volume of the lattice. One sees clearly that both measures agree
for small $\beta$. Increasing $\beta$ stepwise a sudden jump at
a critical $\beta_0$ to positive values occurs for the scale
invariant measure whereas the transition from
negative to positive curvature seems to be smooth in the case of the
uniform measure. We define $\beta_0$ as the transition point from
negative to positive action which can depend on the measure.

It is interesting to investigate separately the behavior of the
areas $A_t$ and of the deficit angles $\delta_t$.
Fig. 2 shows $\langle A_t\rangle^{\frac{1}{2}}$ versus
$\langle q_l\rangle^{\frac{1}{2}}$ and again M1 and M2 yield the
same results for small $\beta$. However, for M2 the areas grow
linearly with the squared link lengths for all $\beta$ values whereas
M1 shows a clear deviation from the linear behavior for
$\beta > \beta_0 $. The reason is the growth of spikes across the
transition point for the invariant measure which will be discussed
below. This gives also rise for a fractal dimension $d<4$ of the
simplicial lattice.

In Fig. 3 both $\langle \delta_t\rangle$ and $\langle \delta_t^2 \rangle$
are drawn versus $\langle q_l\rangle^{\frac{1}{2}}$.
Surprisingly the average of the deficit angles stays negative for all
$\beta$. The change of sign in $S_R$ must be due to a correlation of
a few positive $\delta_t$ with large $A_t$ and many negative $\delta_t$
with small $A_t$.
The value $\langle \delta_t^2 \rangle$ turns out to
decrease in the
small $\beta$ region where M1 and M2 give (nearly) the same results and
shows an increasing behavior in the right phase.

The transition at $\beta_0$ is clearly seen also in the histograms of
single configurations $\{q_l\}$.
In the small $\beta$ region the equilibrium distributions of the
link lengths are rather homogeneous having a small width.
As seen in Fig. 4 a characteristic change of the distributions
occurs for larger $\beta$ leading to inhomogeneous configurations.
For M1 we observe a few very large links and a drastical increase
of the number of short links.
For M2 both the number of small links and the number of links near
the cutoff increases.

To better understand this behavior we calculated the next
neighbor distances $\bar{q}_v$ by averaging $q_l$ over all links meeting
at the same vertex $v$. These distances are plotted in Fig. 5 both for
the low and high $\beta$ region. One observes the formation of isolated
"spikes" in the large $\beta$ region for M1.
The site-to-site fluctuations of $\bar{q}_v$ for M2 indicate a crumpled
lattice without extremely large spikes.

A few remarks about our experience with numerical convergence are in
order.
Using inhomogeneous start configurations it is difficult within M1 to
reach an equilibrium even in the small $\beta$ region and very long
runs are necessary. For M2 the calculated observables converge rather
fast and show no dependence on the start configuration for all $\beta$
values.

To conclude, we have investigated the influence of the measure in the
Regge discretization of Euclidean quantum gravity. Comparing a scale
invariant and a uniform measure it turned out that there exists
a phase of short link lengths in units of the Planck scale where all
observables under consideration were independent of the two measures.
The differences between the expectation values
and the link lengths distribution for larger $\beta$ seem to indicate
an influence of the measure. Investigations of a system with a scale
invariant measure and fixed mean volume yield essentially the same results
as fixed mean link length. At present simulations with a measure
$\prod dq q^{(s-1)} , 0 < s \le 1 , $
interpolating between the uniform and the scale invariant measure
are in progress. Preliminary results show again a phase independent
of the measure indicating a universality of the Regge calculus.

We benefitted from helpful discussions with B. Berg, C. J. Isham and
P. Menotti during the preparation of the article. This work was supported
in part by "Fonds zur F\"orderung der wissenschaftlichen Forschung"
under Contract P7510-TEC.

\begin{center}
\setlength{\unitlength}{0.240900pt}
\ifx\plotpoint\undefined\newsavebox{\plotpoint}\fi
\sbox{\plotpoint}{\rule[-0.175pt]{0.350pt}{0.350pt}}%
\begin{picture}(1125,900)(0,0)
\tenrm
\put(264,368){\rule[-0.175pt]{191.997pt}{0.350pt}}
\put(264,158){\rule[-0.175pt]{0.350pt}{151.526pt}}
\put(264,228){\rule[-0.175pt]{4.818pt}{0.350pt}}
\put(242,228){\makebox(0,0)[r]{-500}}
\put(1041,228){\rule[-0.175pt]{4.818pt}{0.350pt}}
\put(264,368){\rule[-0.175pt]{4.818pt}{0.350pt}}
\put(242,368){\makebox(0,0)[r]{0}}
\put(1041,368){\rule[-0.175pt]{4.818pt}{0.350pt}}
\put(264,507){\rule[-0.175pt]{4.818pt}{0.350pt}}
\put(242,507){\makebox(0,0)[r]{500}}
\put(1041,507){\rule[-0.175pt]{4.818pt}{0.350pt}}
\put(264,647){\rule[-0.175pt]{4.818pt}{0.350pt}}
\put(242,647){\makebox(0,0)[r]{1000}}
\put(1041,647){\rule[-0.175pt]{4.818pt}{0.350pt}}
\put(264,787){\rule[-0.175pt]{4.818pt}{0.350pt}}
\put(242,787){\makebox(0,0)[r]{1500}}
\put(1041,787){\rule[-0.175pt]{4.818pt}{0.350pt}}
\put(264,158){\rule[-0.175pt]{0.350pt}{4.818pt}}
\put(264,113){\makebox(0,0){0}}
\put(264,767){\rule[-0.175pt]{0.350pt}{4.818pt}}
\put(423,158){\rule[-0.175pt]{0.350pt}{4.818pt}}
\put(423,113){\makebox(0,0){0.4}}
\put(423,767){\rule[-0.175pt]{0.350pt}{4.818pt}}
\put(583,158){\rule[-0.175pt]{0.350pt}{4.818pt}}
\put(583,113){\makebox(0,0){0.8}}
\put(583,767){\rule[-0.175pt]{0.350pt}{4.818pt}}
\put(742,158){\rule[-0.175pt]{0.350pt}{4.818pt}}
\put(742,113){\makebox(0,0){1.2}}
\put(742,767){\rule[-0.175pt]{0.350pt}{4.818pt}}
\put(902,158){\rule[-0.175pt]{0.350pt}{4.818pt}}
\put(902,113){\makebox(0,0){1.6}}
\put(902,767){\rule[-0.175pt]{0.350pt}{4.818pt}}
\put(1061,158){\rule[-0.175pt]{0.350pt}{4.818pt}}
\put(1061,113){\makebox(0,0){2}}
\put(1061,767){\rule[-0.175pt]{0.350pt}{4.818pt}}
\put(264,158){\rule[-0.175pt]{191.997pt}{0.350pt}}
\put(1061,158){\rule[-0.175pt]{0.350pt}{151.526pt}}
\put(264,787){\rule[-0.175pt]{191.997pt}{0.350pt}}
\put(45,472){\makebox(0,0)[l]{\shortstack{$\frac{\left<\hbox{S}_E\right>}{\left<\hbox{V}\right>}$}}}
\put(662,58){\makebox(0,0){$\left<\hbox{q}_{\hbox{$l$}}\right>^{1/2}$}}
\put(264,158){\rule[-0.175pt]{0.350pt}{151.526pt}}
\put(344,745){\makebox(0,0)[r]{M1}}
\put(388,745){\raisebox{-1.2pt}{\makebox(0,0){$\Box$}}}
\put(315,202){\raisebox{-1.2pt}{\makebox(0,0){$\Box$}}}
\put(336,288){\raisebox{-1.2pt}{\makebox(0,0){$\Box$}}}
\put(352,316){\raisebox{-1.2pt}{\makebox(0,0){$\Box$}}}
\put(366,330){\raisebox{-1.2pt}{\makebox(0,0){$\Box$}}}
\put(378,339){\raisebox{-1.2pt}{\makebox(0,0){$\Box$}}}
\put(389,344){\raisebox{-1.2pt}{\makebox(0,0){$\Box$}}}
\put(399,349){\raisebox{-1.2pt}{\makebox(0,0){$\Box$}}}
\put(408,352){\raisebox{-1.2pt}{\makebox(0,0){$\Box$}}}
\put(417,354){\raisebox{-1.2pt}{\makebox(0,0){$\Box$}}}
\put(425,356){\raisebox{-1.2pt}{\makebox(0,0){$\Box$}}}
\put(441,420){\raisebox{-1.2pt}{\makebox(0,0){$\Box$}}}
\put(468,482){\raisebox{-1.2pt}{\makebox(0,0){$\Box$}}}
\put(514,490){\raisebox{-1.2pt}{\makebox(0,0){$\Box$}}}
\put(553,496){\raisebox{-1.2pt}{\makebox(0,0){$\Box$}}}
\put(587,517){\raisebox{-1.2pt}{\makebox(0,0){$\Box$}}}
\put(617,537){\raisebox{-1.2pt}{\makebox(0,0){$\Box$}}}
\put(672,594){\raisebox{-1.2pt}{\makebox(0,0){$\Box$}}}
\put(344,700){\makebox(0,0)[r]{M2}}
\put(388,700){\makebox(0,0){$\triangle$}}
\put(314,199){\makebox(0,0){$\triangle$}}
\put(335,285){\makebox(0,0){$\triangle$}}
\put(351,313){\makebox(0,0){$\triangle$}}
\put(377,336){\makebox(0,0){$\triangle$}}
\put(402,347){\makebox(0,0){$\triangle$}}
\put(423,353){\makebox(0,0){$\triangle$}}
\put(459,358){\makebox(0,0){$\triangle$}}
\put(489,361){\makebox(0,0){$\triangle$}}
\put(538,364){\makebox(0,0){$\triangle$}}
\put(581,365){\makebox(0,0){$\triangle$}}
\put(679,368){\makebox(0,0){$\triangle$}}
\put(821,374){\makebox(0,0){$\triangle$}}
\put(998,373){\makebox(0,0){$\triangle$}}
\end{picture}
\\[9cm]
\end{center}
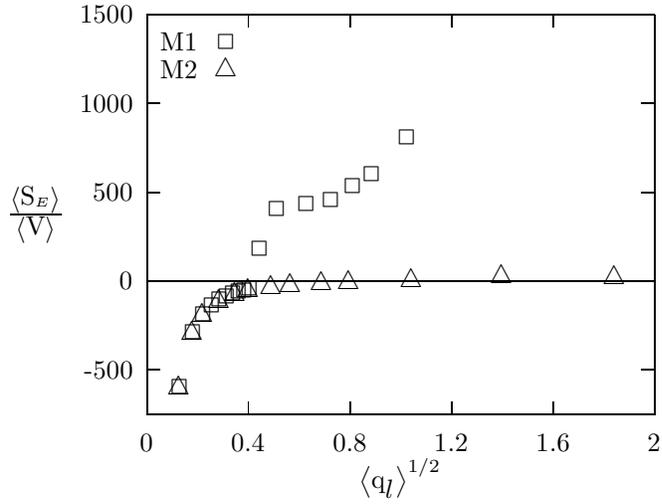
FIG. 1. Action density versus link length in units of $L_P$ for
the scale invariant measure M1 and the uniform measure M2 on
lattice size $4^4$. Averages are taken over the whole lattice and
over 1500 iterations for every $\beta$. Error bars due to mean standard
deviation are smaller than the symbols. In the small $\beta$ region for both
measures a stable regime with negative action density is found.
With increasing $\beta$ M1 exhibits a sudden jump to positive values
whereas M2 shows a smooth behavior.

\newpage

\begin{center}
\setlength{\unitlength}{0.240900pt}
\ifx\plotpoint\undefined\newsavebox{\plotpoint}\fi
\begin{picture}(1125,900)(0,0)
\tenrm
\put(264,158){\rule[-0.175pt]{191.997pt}{0.350pt}}
\put(264,158){\rule[-0.175pt]{0.350pt}{151.526pt}}
\put(264,158){\rule[-0.175pt]{4.818pt}{0.350pt}}
\put(242,158){\makebox(0,0)[r]{0}}
\put(1041,158){\rule[-0.175pt]{4.818pt}{0.350pt}}
\put(264,263){\rule[-0.175pt]{4.818pt}{0.350pt}}
\put(242,263){\makebox(0,0)[r]{0.2}}
\put(1041,263){\rule[-0.175pt]{4.818pt}{0.350pt}}
\put(264,368){\rule[-0.175pt]{4.818pt}{0.350pt}}
\put(242,368){\makebox(0,0)[r]{0.4}}
\put(1041,368){\rule[-0.175pt]{4.818pt}{0.350pt}}
\put(264,473){\rule[-0.175pt]{4.818pt}{0.350pt}}
\put(242,473){\makebox(0,0)[r]{0.6}}
\put(1041,473){\rule[-0.175pt]{4.818pt}{0.350pt}}
\put(264,577){\rule[-0.175pt]{4.818pt}{0.350pt}}
\put(242,577){\makebox(0,0)[r]{0.8}}
\put(1041,577){\rule[-0.175pt]{4.818pt}{0.350pt}}
\put(264,682){\rule[-0.175pt]{4.818pt}{0.350pt}}
\put(242,682){\makebox(0,0)[r]{1}}
\put(1041,682){\rule[-0.175pt]{4.818pt}{0.350pt}}
\put(264,787){\rule[-0.175pt]{4.818pt}{0.350pt}}
\put(242,787){\makebox(0,0)[r]{1.2}}
\put(1041,787){\rule[-0.175pt]{4.818pt}{0.350pt}}
\put(264,158){\rule[-0.175pt]{0.350pt}{4.818pt}}
\put(264,113){\makebox(0,0){0}}
\put(264,767){\rule[-0.175pt]{0.350pt}{4.818pt}}
\put(423,158){\rule[-0.175pt]{0.350pt}{4.818pt}}
\put(423,113){\makebox(0,0){0.4}}
\put(423,767){\rule[-0.175pt]{0.350pt}{4.818pt}}
\put(583,158){\rule[-0.175pt]{0.350pt}{4.818pt}}
\put(583,113){\makebox(0,0){0.8}}
\put(583,767){\rule[-0.175pt]{0.350pt}{4.818pt}}
\put(742,158){\rule[-0.175pt]{0.350pt}{4.818pt}}
\put(742,113){\makebox(0,0){1.2}}
\put(742,767){\rule[-0.175pt]{0.350pt}{4.818pt}}
\put(902,158){\rule[-0.175pt]{0.350pt}{4.818pt}}
\put(902,113){\makebox(0,0){1.6}}
\put(902,767){\rule[-0.175pt]{0.350pt}{4.818pt}}
\put(1061,158){\rule[-0.175pt]{0.350pt}{4.818pt}}
\put(1061,113){\makebox(0,0){2}}
\put(1061,767){\rule[-0.175pt]{0.350pt}{4.818pt}}
\put(264,158){\rule[-0.175pt]{191.997pt}{0.350pt}}
\put(1061,158){\rule[-0.175pt]{0.350pt}{151.526pt}}
\put(264,787){\rule[-0.175pt]{191.997pt}{0.350pt}}
\put(45,472){\makebox(0,0)[l]{\shortstack{$\left<A_t\right>^{1/2}$}}}
\put(662,58){\makebox(0,0){$\left<\hbox{q}_{\hbox{$l$}}\right>^{1/2}$}}
\put(264,158){\rule[-0.175pt]{0.350pt}{151.526pt}}
\put(344,745){\makebox(0,0)[r]{M1}}
\put(388,745){\raisebox{-1.2pt}{\makebox(0,0){$\Box$}}}
\put(315,200){\raisebox{-1.2pt}{\makebox(0,0){$\Box$}}}
\put(336,217){\raisebox{-1.2pt}{\makebox(0,0){$\Box$}}}
\put(352,230){\raisebox{-1.2pt}{\makebox(0,0){$\Box$}}}
\put(366,241){\raisebox{-1.2pt}{\makebox(0,0){$\Box$}}}
\put(378,251){\raisebox{-1.2pt}{\makebox(0,0){$\Box$}}}
\put(389,260){\raisebox{-1.2pt}{\makebox(0,0){$\Box$}}}
\put(399,268){\raisebox{-1.2pt}{\makebox(0,0){$\Box$}}}
\put(408,276){\raisebox{-1.2pt}{\makebox(0,0){$\Box$}}}
\put(417,283){\raisebox{-1.2pt}{\makebox(0,0){$\Box$}}}
\put(425,289){\raisebox{-1.2pt}{\makebox(0,0){$\Box$}}}
\put(441,252){\raisebox{-1.2pt}{\makebox(0,0){$\Box$}}}
\put(468,261){\raisebox{-1.2pt}{\makebox(0,0){$\Box$}}}
\put(514,289){\raisebox{-1.2pt}{\makebox(0,0){$\Box$}}}
\put(553,304){\raisebox{-1.2pt}{\makebox(0,0){$\Box$}}}
\put(587,334){\raisebox{-1.2pt}{\makebox(0,0){$\Box$}}}
\put(617,354){\raisebox{-1.2pt}{\makebox(0,0){$\Box$}}}
\put(672,378){\raisebox{-1.2pt}{\makebox(0,0){$\Box$}}}
\put(344,700){\makebox(0,0)[r]{M2}}
\put(388,700){\makebox(0,0){$\triangle$}}
\put(314,200){\makebox(0,0){$\triangle$}}
\put(335,218){\makebox(0,0){$\triangle$}}
\put(351,231){\makebox(0,0){$\triangle$}}
\put(377,252){\makebox(0,0){$\triangle$}}
\put(402,273){\makebox(0,0){$\triangle$}}
\put(423,291){\makebox(0,0){$\triangle$}}
\put(459,320){\makebox(0,0){$\triangle$}}
\put(489,345){\makebox(0,0){$\triangle$}}
\put(538,386){\makebox(0,0){$\triangle$}}
\put(581,421){\makebox(0,0){$\triangle$}}
\put(679,499){\makebox(0,0){$\triangle$}}
\put(821,608){\makebox(0,0){$\triangle$}}
\put(998,748){\makebox(0,0){$\triangle$}}
\end{picture}
\\[10cm]
\end{center}
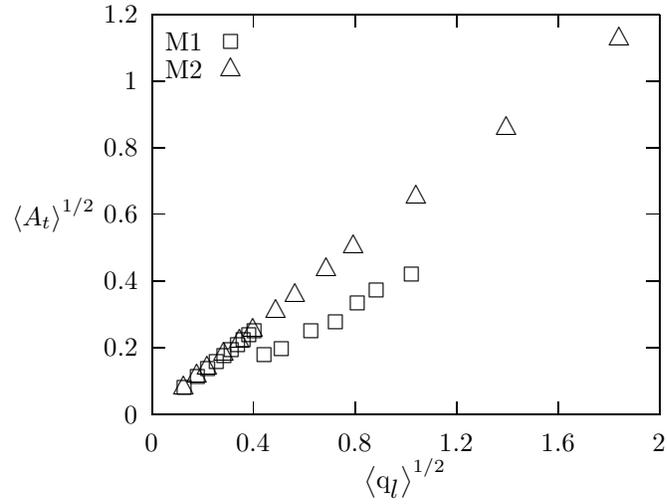
FIG. 2. Square root of average area versus average link length
in units of $L_P$ computed for the scale invariant (M1) and the uniform
(M2) measure.  M2 shows a linear increase in the entire $\beta$ regime
whereas M1 exhibits a sudden deviation from the linear behavior at $\beta_0$.

\newpage

\begin{center}
\setlength{\unitlength}{0.240900pt}
\ifx\plotpoint\undefined\newsavebox{\plotpoint}\fi
\begin{picture}(1125,900)(0,0)
\tenrm
\put(264,368){\rule[-0.175pt]{191.997pt}{0.350pt}}
\put(264,158){\rule[-0.175pt]{0.350pt}{151.526pt}}
\put(264,158){\rule[-0.175pt]{4.818pt}{0.350pt}}
\put(242,158){\makebox(0,0)[r]{-2}}
\put(1041,158){\rule[-0.175pt]{4.818pt}{0.350pt}}
\put(264,263){\rule[-0.175pt]{4.818pt}{0.350pt}}
\put(242,263){\makebox(0,0)[r]{-1}}
\put(1041,263){\rule[-0.175pt]{4.818pt}{0.350pt}}
\put(264,368){\rule[-0.175pt]{4.818pt}{0.350pt}}
\put(242,368){\makebox(0,0)[r]{0}}
\put(1041,368){\rule[-0.175pt]{4.818pt}{0.350pt}}
\put(264,473){\rule[-0.175pt]{4.818pt}{0.350pt}}
\put(242,473){\makebox(0,0)[r]{5}}
\put(1041,473){\rule[-0.175pt]{4.818pt}{0.350pt}}
\put(264,577){\rule[-0.175pt]{4.818pt}{0.350pt}}
\put(242,577){\makebox(0,0)[r]{10}}
\put(1041,577){\rule[-0.175pt]{4.818pt}{0.350pt}}
\put(264,682){\rule[-0.175pt]{4.818pt}{0.350pt}}
\put(242,682){\makebox(0,0)[r]{15}}
\put(1041,682){\rule[-0.175pt]{4.818pt}{0.350pt}}
\put(264,787){\rule[-0.175pt]{4.818pt}{0.350pt}}
\put(242,787){\makebox(0,0)[r]{20}}
\put(1041,787){\rule[-0.175pt]{4.818pt}{0.350pt}}
\put(264,158){\rule[-0.175pt]{0.350pt}{4.818pt}}
\put(264,113){\makebox(0,0){0}}
\put(264,767){\rule[-0.175pt]{0.350pt}{4.818pt}}
\put(423,158){\rule[-0.175pt]{0.350pt}{4.818pt}}
\put(423,113){\makebox(0,0){0.4}}
\put(423,767){\rule[-0.175pt]{0.350pt}{4.818pt}}
\put(583,158){\rule[-0.175pt]{0.350pt}{4.818pt}}
\put(583,113){\makebox(0,0){0.8}}
\put(583,767){\rule[-0.175pt]{0.350pt}{4.818pt}}
\put(742,158){\rule[-0.175pt]{0.350pt}{4.818pt}}
\put(742,113){\makebox(0,0){1.2}}
\put(742,767){\rule[-0.175pt]{0.350pt}{4.818pt}}
\put(902,158){\rule[-0.175pt]{0.350pt}{4.818pt}}
\put(902,113){\makebox(0,0){1.6}}
\put(902,767){\rule[-0.175pt]{0.350pt}{4.818pt}}
\put(1061,158){\rule[-0.175pt]{0.350pt}{4.818pt}}
\put(1061,113){\makebox(0,0){2}}
\put(1061,767){\rule[-0.175pt]{0.350pt}{4.818pt}}
\put(264,158){\rule[-0.175pt]{191.997pt}{0.350pt}}
\put(1061,158){\rule[-0.175pt]{0.350pt}{151.526pt}}
\put(264,787){\rule[-0.175pt]{191.997pt}{0.350pt}}
\put(105,263){\makebox(0,0)[l]{\shortstack{$\left<\delta_t\right>$}}}
\put(105,577){\makebox(0,0)[l]{\shortstack{$\left<\delta_t^2\right>$}}}
\put(662,58){\makebox(0,0){$\left<\hbox{q}_{\hbox{$l$}}\right>^{1/2}$}}
\put(264,158){\rule[-0.175pt]{0.350pt}{151.526pt}}
\put(344,745){\makebox(0,0)[r]{M1}}
\put(388,745){\raisebox{-1.2pt}{\makebox(0,0){$\Box$}}}
\put(315,439){\raisebox{-1.2pt}{\makebox(0,0){$\Box$}}}
\put(336,438){\raisebox{-1.2pt}{\makebox(0,0){$\Box$}}}
\put(352,437){\raisebox{-1.2pt}{\makebox(0,0){$\Box$}}}
\put(366,437){\raisebox{-1.2pt}{\makebox(0,0){$\Box$}}}
\put(378,436){\raisebox{-1.2pt}{\makebox(0,0){$\Box$}}}
\put(389,435){\raisebox{-1.2pt}{\makebox(0,0){$\Box$}}}
\put(399,434){\raisebox{-1.2pt}{\makebox(0,0){$\Box$}}}
\put(408,434){\raisebox{-1.2pt}{\makebox(0,0){$\Box$}}}
\put(417,433){\raisebox{-1.2pt}{\makebox(0,0){$\Box$}}}
\put(425,433){\raisebox{-1.2pt}{\makebox(0,0){$\Box$}}}
\put(441,438){\raisebox{-1.2pt}{\makebox(0,0){$\Box$}}}
\put(468,440){\raisebox{-1.2pt}{\makebox(0,0){$\Box$}}}
\put(514,440){\raisebox{-1.2pt}{\makebox(0,0){$\Box$}}}
\put(553,440){\raisebox{-1.2pt}{\makebox(0,0){$\Box$}}}
\put(587,446){\raisebox{-1.2pt}{\makebox(0,0){$\Box$}}}
\put(617,451){\raisebox{-1.2pt}{\makebox(0,0){$\Box$}}}
\put(672,454){\raisebox{-1.2pt}{\makebox(0,0){$\Box$}}}
\put(344,700){\makebox(0,0)[r]{M2}}
\put(388,700){\makebox(0,0){$\triangle$}}
\put(314,432){\makebox(0,0){$\triangle$}}
\put(335,431){\makebox(0,0){$\triangle$}}
\put(351,431){\makebox(0,0){$\triangle$}}
\put(377,429){\makebox(0,0){$\triangle$}}
\put(402,427){\makebox(0,0){$\triangle$}}
\put(423,425){\makebox(0,0){$\triangle$}}
\put(459,421){\makebox(0,0){$\triangle$}}
\put(489,418){\makebox(0,0){$\triangle$}}
\put(538,413){\makebox(0,0){$\triangle$}}
\put(581,409){\makebox(0,0){$\triangle$}}
\put(679,416){\makebox(0,0){$\triangle$}}
\put(821,591){\makebox(0,0){$\triangle$}}
\put(998,736){\makebox(0,0){$\triangle$}}
\put(315,366){\raisebox{-1.2pt}{\makebox(0,0){$\Box$}}}
\put(336,366){\raisebox{-1.2pt}{\makebox(0,0){$\Box$}}}
\put(352,366){\raisebox{-1.2pt}{\makebox(0,0){$\Box$}}}
\put(366,366){\raisebox{-1.2pt}{\makebox(0,0){$\Box$}}}
\put(378,365){\raisebox{-1.2pt}{\makebox(0,0){$\Box$}}}
\put(389,365){\raisebox{-1.2pt}{\makebox(0,0){$\Box$}}}
\put(399,365){\raisebox{-1.2pt}{\makebox(0,0){$\Box$}}}
\put(408,365){\raisebox{-1.2pt}{\makebox(0,0){$\Box$}}}
\put(417,365){\raisebox{-1.2pt}{\makebox(0,0){$\Box$}}}
\put(425,364){\raisebox{-1.2pt}{\makebox(0,0){$\Box$}}}
\put(441,364){\raisebox{-1.2pt}{\makebox(0,0){$\Box$}}}
\put(468,362){\raisebox{-1.2pt}{\makebox(0,0){$\Box$}}}
\put(514,360){\raisebox{-1.2pt}{\makebox(0,0){$\Box$}}}
\put(553,359){\raisebox{-1.2pt}{\makebox(0,0){$\Box$}}}
\put(587,353){\raisebox{-1.2pt}{\makebox(0,0){$\Box$}}}
\put(617,349){\raisebox{-1.2pt}{\makebox(0,0){$\Box$}}}
\put(672,346){\raisebox{-1.2pt}{\makebox(0,0){$\Box$}}}
\put(314,367){\makebox(0,0){$\triangle$}}
\put(335,367){\makebox(0,0){$\triangle$}}
\put(351,367){\makebox(0,0){$\triangle$}}
\put(377,367){\makebox(0,0){$\triangle$}}
\put(402,367){\makebox(0,0){$\triangle$}}
\put(423,367){\makebox(0,0){$\triangle$}}
\put(459,367){\makebox(0,0){$\triangle$}}
\put(489,367){\makebox(0,0){$\triangle$}}
\put(538,367){\makebox(0,0){$\triangle$}}
\put(581,366){\makebox(0,0){$\triangle$}}
\put(679,357){\makebox(0,0){$\triangle$}}
\put(821,283){\makebox(0,0){$\triangle$}}
\put(998,242){\makebox(0,0){$\triangle$}}
\end{picture}
\\[10cm]
\end{center}
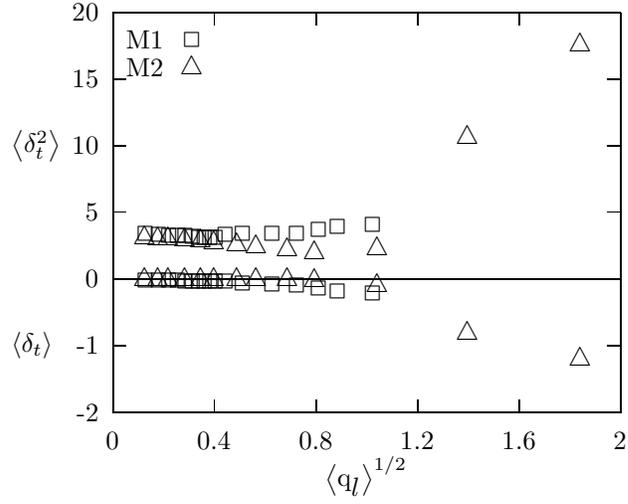
FIG. 3. Average deficit angle and squared deficit angle
versus link length. $\langle \delta_t \rangle$ stays
negative even for $\beta$'s with positive curvature and
$\langle \delta_t^2 \rangle$ reaches its minimum at the transition
point $\beta_0$.

\newpage

\begin{center}
\setlength{\unitlength}{0.240900pt}
\ifx\plotpoint\undefined\newsavebox{\plotpoint}\fi
\sbox{\plotpoint}{\rule[-0.175pt]{0.350pt}{0.350pt}}%
\begin{picture}(1125,900)(0,0)
\end{picture}
\\[10cm]
\end{center}
FIG. 4. Histograms of squared link lengths in units of the average link
length in double logarithmic scale. The shapes
are nearly independent of the measure for low $\beta$. For
higher $\beta$ the histograms differ significantly. In M1 the formation
of a few large and a lot of small links occurs while for M2 the lack of
small links is a characteristics of the measure.

\newpage

\begin{center}
\setlength{\unitlength}{0.240900pt}
\ifx\plotpoint\undefined\newsavebox{\plotpoint}\fi
\sbox{\plotpoint}{\rule[-0.175pt]{0.350pt}{0.350pt}}%
\begin{picture}(1125,900)(0,0)
\end{picture}
\\[10cm]
\end{center}
FIG. 5. Average squared distance $\bar{q}_v$ of the vertices from their
next neighbors in units of the average link length. The pictures
correspond to the configurations of Fig. 4. For higher $\beta$ the scale
invariant measure develops spikes whereas a crumpled lattice without
significant spikes is observed for the uniform measure.

\end{document}